\documentclass[%
 reprint,
 amsmath,amssymb,
pra,
]{revtex4-1}

\usepackage{graphicx}
\usepackage{dcolumn}
\usepackage{bm}
\usepackage{indentfirst}
\usepackage[utf8]{inputenc}
\usepackage{epstopdf}

\begin{document}

\preprint{APS/123-QED}

\title{Path integral approach to the problem of rotational excitation of molecules \\ by an ultrashort laser pulses  sequence }

\author{Alexander Biryukov}
\email{biryukov@samsu.ru}
\author{Mark Shleenkov}%
 \email{shleenkov@list.ru}
\affiliation{%
Samara State University, Pavlov Street 1, Samara 443011, Russia}%

\date{\today}

\begin{abstract}
The amplitude and probability of quantum transitions are represented as a path integrals in energy state space of the investigated multi-level quantum system. Using this approach we consider rotational dynamics of nitrogen molecules $^{14}N_2$ and $^{15}N_{2}$ which interact with a sequence of ultrashort laser pulses. Our computer simulations indicate the complex dependency of the high rotation states excitation probability upon ultrashort laser pulses sequence periods. We observe pronounced resonances, which correspond to the results of some experiments.
\begin{description}
\item[PACS numbers]
33.80.-b, 42.65.Re, 31.15.xk.
\end{description}
\end{abstract}

\maketitle

\section{\label{sec:level1} Introduction}

The modern development of laser radiation technologies induces theoretical and experimental investigations of the dynamics of quantum objects (such as atoms or molecules) under the action of intense electromagnetic field of different forms.

This dynamics is principally non-linear, because the probability is high of multiphoton processes (absorption and emission more the one photon) and nonresonant processes (electromagnetic field frequency is far from quantum transitions frequency). We note the recent studies of different rare gases multiphoton ionization \cite{gerken14, guichard13, richter09}, of multiphoton photoemission of the Au(111) surface state with 800-nm laser pulses \cite{sirotti14}, of multiphoton transitions in GaSb/GaAs quantum-dot intermediate-band solar cells \cite{hwang14}, of three-photon electromagnetically induced absorption in a ladder-type atomic system \cite{moon14}.

There are certain difficulties for theoretical studies of these processes and for simulations of quantum objects dynamics that interact with laser field. Thus, different approximations are used. For example, there are two- or three-level quantum system models \cite{cho14} and rotating wave approximation \cite{spiegelberg13}.
For high-intensity laser field the perturbation theory runs into problems. It is necessary to calculate the large number of terms. High-order perturbation theory for miltilevel quantum system dynamics was considered in \cite{biryukov08}.
For theoretical researches of this processes the numerical solution of time-dependent Schr\"odinger equation is used \cite{fleischer09}. For this reason different schemes of space-time discretization is realized. The discretization parameter should be small enough for simulations of a minute error.   

The perspective approach to theoretical studies of this quantum processes is path integral (functional integral) formalism, which are formulated by R.P. Feynman \cite{feyn48, feyn65} and based on P.A.M. Dirac ideas \cite{dirac33, dirac82}. At present this formalism is an abundantly used approach in many fields of physics: lattice theories in QCD simulations \cite{bornyakov13} and those in graphene \cite{pavlovsly13}, semiclassical approachs in atom optics \cite{bichkov09, bichkov11}, black-swan events problem \cite{kleinert13}, influence functional approach in quantum theory \cite{feyn63, biryukov13} and many others.

In this paper we present a new theoretical approach for describing the dynamics of a quantum system, interacting with laser radiation by path integration in energy states space. We obtain formulas for calculating the quantum transition amplitude and probability as path integrals in energy states space (the space of discrete non-negative variables i.e. quantum numbers).

Recent experimental \cite{zhdanovich12} and theoretical \cite{floss12, floss13} investigations point at possibilities of selective excitations of nitrogen isotopes by a sequence of ultrashort laser pulses (a pulse train). We have developed and are applying the theoretical approach to quantum resonances problem in molecule rotational excitation by ultrashort laser pulses.

\section{Quantum transition probability as path integral in energy states space}

We consider interaction of multilevel quantum system (such as an atom or a molecule) with electromagnetic field. The Hamiltonian $\hat{H}_{full}$ describing our model is given as
\begin{equation}
\hat{H}_{full}=\hat{H}_{syst}+\hat{V},
\label{eq:1}
\end{equation}
where $\hat{H}_{syst}$ is Hamiltonian of the investigated quantum system. We define stationary eigenstates $|l\rangle$ with energies $E_l$ having the following properties:
\begin{eqnarray}
\hat{H}_{syst}=\sum\limits_{l=0}^{N-1} E_{l} |l\rangle \langle l|, \hspace{0.6cm}
\label{eq:2}
\\
\sum\limits_{l=0}^{N-1} |l\rangle \langle l|=1, \ \ \ \ \ \ \langle l' | l \rangle = \delta_{l'l};
\label{eq:3}
\end{eqnarray}
$\hat{V}$ -- the interaction operator.

Our main goal is to define the probability $P(l_{f},t|l_{in},0)$ of investigated quantum system transition from eigenstate $|l_{in}\rangle$ at the moment $t=0$ to the one $|l_{f}\rangle$ at the moment $t>0$.

We describe the investigated system by statistical operator $\hat{\rho}(t)$. The evolution equation of $\hat{\rho}(t)$ in Dirac (interaction) picture \cite{dirac82} is as follows:
\begin{equation}
\hat{\rho}(t)=\hat{U}_D(t)\hat{\rho}(0)\hat{U}_D^+(t),
\label{eq:4}
\end{equation}
where
\begin{equation}
\hat{U}_{D}(t)=T\exp[-\frac{\imath}{\hbar}\int\limits_{0}^t \hat{V}_D(\tau) d\tau]
\label{eq:5}
\end{equation}
--  the evolution operator in Dirac picture,
\begin{equation}
\hat{V}_D(\tau)=\exp[\frac{\imath}{\hbar} \hat{H}_{syst}\tau] \hat{V}(\tau) \exp[-\frac{\imath}{\hbar} \hat{H}_{syst}\tau]
\label{eq:6}
\end{equation}
-- the operator of quantum system and electromagnetic field interaction in Dirac picture.

Eq.~(\ref{eq:4}) in energy representation is
\begin{equation}
\rho_{l_f m_f}(t)=\sum\limits_{l_{in},m_{in}}\langle l_{f}|\hat{U}_{D}(t)|l_{in}\rangle \rho_{l_{in}m_{in}} \langle m_{in} | \hat{U}_{D}^+(t)|m_{f} \rangle,%
\label{eq:7}
\end{equation}
where 
\begin{equation}
\rho_{l_{f} m_{f}}(t)=\langle l_{f}| \hat{\rho}(t) | m_{f} \rangle, \ \ \ \
\rho_{l_{in},m_{in}}=\langle l_{in}| \hat{\rho}(0) | m_{in}' \rangle.
\nonumber
\end{equation}

By the use of evolution operator $\hat{U}$ group properties and completeness condition Eq.~(\ref{eq:3}) of eigenvectors $|l_k\rangle$ basis the kernel of evolution operator $\langle l_f| \hat{U}_{D}(t)| l_{in}\rangle$ can be expressed as
\begin{equation}
\langle l_f| \hat{U}_{D}(t)| l_{in}\rangle =
\sum\limits_{l_{1},..,l_{K}=0}^{N-1}\prod\limits_{k=1}^{K+1}\langle l_k| \hat{U}_{D}(t_{k},t_{k-1})| l_{k-1}\rangle,
\label{eq:8}
\end{equation}
as long as $t_{k}>t_{k-1}$ and where
\begin{equation}
\hat{U}_D(t_{k},t_{k-1})=\exp[-\frac{\imath}{\hbar}\int\limits_{t_{k-1}}^{t_{k}}
\hat{V}_D(\tau) d\tau],
\label{eq:9}
\end{equation}
here we introduce the notations $t_{K+1}=t$, $l_{K+1}=l_f$, $t_{0}=0$, $l_{0}=l_{in}$, $\sum\limits_{k=1}^{K+1}(t_{k}-t_{k-1})=t$.

In Appendix A we show that for small time interval $(t_k-t_{k-1})\rightarrow 0$ the evolution operator kernel $\langle l_k| \hat{U}_D(t_{k},t_{k-1})| l_{k-1}\rangle$ can be expressed as
\begin{equation}
\langle l_k| \hat{U}_{D}(t_{k},t_{k-1})| l_{k-1}\rangle= \int\limits_{0}^{1} \exp[\imath S[l_k,l_{k-1};\xi_{k-1}]]d\xi_{k-1},
\label{eq:10}
\end{equation}
where $S[l_{k},l_{k-1}\xi_{k-1}]$ -- dimensionless (in $\hbar$ units) action in energy representation during time interval $(t_{k}-t_{k-1})$
\begin{widetext}
\begin{equation}
S[l_k,l_{k-1};\xi_{k-1}]=
2\pi(l_{k}-l_{k-1})\xi_{k-1}-
\int\limits_{t_{k-1}}^{t_{k}} \frac{V_{l_{k}l_{k-1}}(\tau)}{\hbar}
2\cos[2\pi (l_{k}-l_{k-1})\xi_{k-1} - \omega_{l_{k}l_{k-1}}\tau] d\tau,
\label{eq:11}
\end{equation}
\end{widetext}
where $V_{l_{k}l_{k-1}}(\tau)=\langle l_{k}| \hat{V}(\tau) |l_{k-1}\rangle$ -- interaction operator matrix element.

The probability $P(l_{f},t|l_{in},0)$ of transition from pure quantum state $\hat{\rho}(0)=| l_{in} \rangle \langle l_{in}|$ $ (\rho_{l_{in}m_{in}}(0)=\delta_{l_{in}m_{in}})$ at the initial moment $t=0$ to the quantum state $\hat{\rho}(t)=|l_{f}\rangle \langle l_{f}|$ $(\rho_{l_{f}m_{f}}(t)=P_{l_{f}}(t)=\delta_{l_{f}m_{f}})$ at the final moment $t$ can be defined by using Eq.~(\ref{eq:7})
\begin{equation}
P(l_{f},t|l_{in},0)=U_{D}^*(l_{f},t|l_{in},0)U_{D}(l_{f},t|l_{in},0),
\label{eq:12}
\end{equation}
where $U_{D}(l_{f},t|l_{in},0)$ is transition amplitude Eq.~(\ref{eq:8}).

If at the initial moment $t=0$ the state of the quantum system under investigation is expressed as distribution $\hat{\rho}(0) = \sum\limits_{l_{in}=0}^{N-1} P_{l_{in}}(0) |l_{in} \rangle \langle l_{in}|$ ($ \rho_{l_{in}m_{in}}(0)=P_{l_{in}}(0)\delta_{l_{in}m_{in}} $) over the pure eigenstates $|l_{in}\rangle$, the probability of quantum system observation in eigenstate $l_{f}$ at moment $t$ has the following form:
\begin{equation}
P(l_{f},t|\rho(0))=\sum\limits_{l_{in}=0}^{N-1} P(l_{f},t|l_{in},0) P_{l_{in}}(0).
\label{eq:13}
\end{equation}

We note that using Eq.~(\ref{eq:8}), Eq.~(\ref{eq:10}), Eq.~(\ref{eq:11}) quantum transition amplitude $U_D(l_{f},t|l_{in},0)$ for any $t$ can be expressed as path integral in energy eigenstates space
\begin{widetext}
\begin{equation}
\langle l_f| \hat{U}_D(t)| l_{in}\rangle = U_D(l_{f},t|l_{in},0)
=
\lim\limits_{K \rightarrow \infty}\sum\limits_{l_{1},..,l_{K}=0}^{N-1} \int\limits_{0}^{1} ..\int\limits_{0}^{1} \exp[\imath S[l_f,l_{K},\xi_K;..;l_k,l_{k-1},\xi_{k-1};..;l_{1},l_{in},\xi_0]]d\xi_0..d\xi_K,
\label{eq:14}
\end{equation}
\end{widetext}
where
\begin{eqnarray}
S[l_f,l_{K},\xi_K;..; l_k,l_{k-1},\xi_{k-1};..;l_{1},l_{in},\xi_0]= \nonumber\\ =\sum\limits_{k=1}^{K+1}S[l_k,l_{k-1},\xi_{k-1}]
\label{eq:15}
\end{eqnarray}
-- dimensionless action. It is a functional, which is defined on a path set in discrete variables $l_k$ space of size $N$ (quantum system levels number) and continuous c-number variables $\xi_k$ space $[0,1]$.

Using Eq.~(\ref{eq:14}) we can express the density matrix Eq.~(\ref{eq:7}) and the transition probability Eq.~(\ref{eq:12}) as the path integrals:
\begin{widetext}
\begin{eqnarray}
\rho_{l_f,m_f}(t)=\lim\limits_{K \rightarrow \infty} \sum\limits_{l_{in},..,l_{K}=0}^{N-1}\sum\limits_{m_{in},..,m_{K}=0}^{N-1} \int\limits_{0}^{1} ..\int\limits_{0}^{1} d\xi_0..d\xi_Kd\zeta_0..d\zeta_K
\exp[\imath (S[l_f,l_{K},\xi_K;..;l_k,l_{k-1},\xi_{k-1};..;l_{1},l_{in},\xi_0] - \nonumber
\\
-S[m_f,m_{K},\zeta_K;..;m_k,m_{k-1},\zeta_{k-1};..;m_{1},m_{in},\zeta_0])]\rho_{l_{in},m_{in}}(0), \hspace{0.5cm}
\label{eq:16}
\end{eqnarray}
\begin{eqnarray}
P(l_f,t|l_{in},0)=\lim\limits_{K \rightarrow \infty}
\sum\limits_{l_{1},..,l_{K}=0}^{N-1}\sum\limits_{m_{1},..,m_{K}=0}^{N-1} \int\limits_{0}^{1} ..\int\limits_{0}^{1} d\xi_0..d\xi_Kd\zeta_0..d\zeta_K
\exp[\imath (S[l_f,l_{K},\xi_K;..;l_k,l_{k-1},\xi_{k-1};..;l_{1},l_{in},\xi_0] - \nonumber
\\
-S[l_f,m_{K},\zeta_K;..;m_k,m_{k-1},\zeta_{k-1};..;m_{1},l_{in},\zeta_0])].\hspace{0.5cm}
\label{eq:17}
\end{eqnarray}
\end{widetext}

Thus, Eq.~(\ref{eq:16}), Eq.~(\ref{eq:17}) with Eq.~(\ref{eq:15}) and Eq.~(\ref{eq:11}) are the closed equations system for describing of transitions of miltilevel quantum system interacting with electromagnetic field by interaction operator $\hat{V}$.


\section{\label{sec:level2} Rotational dynamics of $\bf{^{14}N_2}$ and $\bf{^{15}N_2}$ interacting  with laser pulses sequences}

Recent results of experimental observation of $^{14}N_2$ and $^{15}N_2$ high rotational states excitation were published in \cite{zhdanovich12}. Detailed discussions of the results were in \cite{floss13, floss12}.

In the experiments the groups of $^{14}N_2$ and $^{15}N_2$ molecules were investigated. At the initial moment the distribution of rotational population is thermal and corresponds to $T=6.3$ K. Molecules interact with a sequence of ultrashort laser pulses with period from $6.5$ ps to $9.5$ ps. Each laser pulse has duration equal $500$ fs. Laser radiation intensity reaches the value $I=5*10^{12}$ W/cm$^2$. The relative populations were measured  of the rotational levels of $^{14}N_2$ and $^{15}N_2$ and the functional dependence  of the populations on the pulse train period was obtained. 

The results of these experiments show that there are quantum nonlinear resonances i.e. the nonlinear increase of rotational excitation efficiency under specific values of the pulse train period. The most efficient population transfer up the rotational ladder occurs around $8.4$ ps for $^{14}N_2$ and $9$ ps for $^{15}N_2$.

We analyse these experiments using the method developed by us which is based on path integral formulation in energy states space.

The initial distribution of rotational population is thermal and corresponds to $T=6.3$ K:

\begin{equation}
P_{l_{in}}=\frac{1}{Z}\exp[-\frac{E_{l_{in}}}{k_{B}T}],
\label{eq:18}
\end{equation}
where
\begin{equation}
Z=\sum\limits_{l_{in}=0}^{N-1}\exp[-\frac{E_{l_{in}}}{k_{B}T}]
\label{eq:19}
\end{equation}
-- particle function, $k$ -- Boltzmann factor, $T$ -- absolute temperature, $N$ -- rotational states number in the theoretical model.

We calculate the energy $E_{l}$ of investigated molecules rotational levels for quantum rigid rotor model \cite{landau76}
\begin{equation}
-\frac{\hbar^2}{2I}\frac{1}{\sin{\theta}}\frac{\partial}{\partial \theta}(\sin{\theta}\frac{\partial}{\partial \theta})Y_l(\theta)=E_{l}Y_l(\theta),
\label{eq:20}
\end{equation}
where $I=\mu R^2$ -- moment of inertia, $\mu$ -- molecule reduced mass, $R$ -- atom distances, $Y_l(\theta)=Y_{l}^{0}(\theta,\phi)$, where $Y_l^m(\theta,\phi)$ -- spherical harmonics.

Eq.~(\ref{eq:20}) defines the rotational energy spectrum of a diatomic molecule
\begin{equation*}
E_l= \frac{\hbar^2}{2I}l(l+1),
\end{equation*}
where $l$ -- azimuthal quantum number.

It is known, that nonpolar molecule dipole moment is equal to zero. However, strong laser fields induce the molecular dipole by exerting an angle-dependent torque. 

The interaction is described by the potential \cite{zon75, underwood05}
\begin{equation}
V(\tau)=-\frac{1}{4}\Delta\alpha E^2(\tau) \cos^2\theta,
\label{eq:21}
\end{equation}
where $\Delta \alpha$ describes the molecular polarizability, $\theta$ is the angle between the molecular axis and the field polarization.
\begin{figure*}
\includegraphics[scale=0.8]{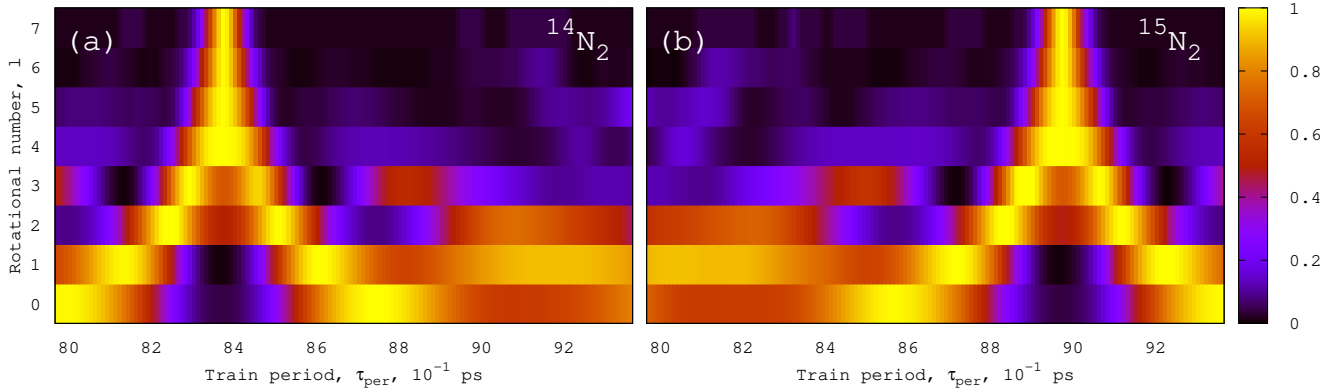}
\caption{\label{fig:1} (color online) Nitrogen $^{14}N_2$ (a) and $^{15}N_2$ (b) molecules observation probabilities in rotation states $l=0,1,\dots 7$ as a function of the rotational number $l$ and the pulses train period  $\tau$. We normalize them by their maximum values for each rotational state. }
\end{figure*}
Matrix elements of interaction operator are
\begin{equation}
V_{l'l}(\tau)=-\frac{1}{4}\Delta\alpha E^2(\tau)\langle l'|\cos^2\theta |l \rangle,
\label{eq:22}
\end{equation}
where
\begin{equation}
\langle l'| \cos^2\theta |l\rangle = 2\pi\int\limits_{0}^{\pi} Y_{l'}^{*}(\theta) \cos^2\theta Y_l(\theta) \sin\theta d\theta.
\label{eq:23}
\end{equation}

Matrix elements $\langle l'| \cos^2\theta |l\rangle$ were numerically calculated by Eq.~(\ref{eq:22}) and Eq.~(\ref{eq:23}).

The investigated molecules parameters are \cite{irikura07}:
$\Delta \alpha = 1.97*10^{-40}$ C$*$m$^2$/V, $I = 1.4*10^{-46}$ kg$*$m$^2$ for $^{14}N_2$, $I = 1.5*10^{-46}$ kg$*$m$^2$ for $^{15}N_2$.

We consider a sequence of ultrashort laser pulses which was used in  \cite{zhdanovich12}. The electric field value is as follows
\begin{equation}
E(\tau)=\sum\limits_{n=-3}^{3} J_{n}(A) E_{0}\exp[-\frac{(\tau-n\tau_{per})^2}{\tau_{pul}^2}],
\label{eq:24}
\end{equation}
were $J_{n}(A)$ is Bessel function of the first kind, $A=2.5$ is the spectral phase modulation amplitude,  $E_0 \approx 6 \times 10^{9}$ V/m is electric field value, $\tau_{pul}\approx 500$ fs is each laser pulse duration, $7.98$ ps $\leq \tau_{per} \leq 9.38$ ps is pulse train period.

We are considering the model of $N_2$ with $N=8$ rotational levels ($l=0,1,\dots,7$). This model is a good approximation, because in experiments \cite{zhdanovich12} higher rotational states are practically not excited.

By the use of Eq.~(\ref{eq:20})-(\ref{eq:24}) and numerical simulation algorithm (see Appendix B) we calculate the probability of excitation from the initial state (Boltzmann distribution) to different rotational states having  interacted with a sequence of ultrashort laser pulses as a function of pulse train period. The absolute error of our probability calculation was not more than  $10^{-3}$.  The results are given in Fig.~\ref{fig:1}.

In Fig.~\ref{fig:1} (two-dimensional map) we present normalized probability of $^{14}N_2$ and $^{15}N_2$ molecules rotational state observation after they have interacted with 7 laser pulses under different pulses train periods. For the pulse train period equal to $8.38$ ps for $^{14}N_2$ and $8.98$ ps for $^{15}N_2$ the population is efficiently transferred from the initial (thermal distribution) states $l=0,1,2$ to the higher states $l=3,4,5,6,7$.


\section{Conclusion}
In this paper we present new method of calculating the transition probability of a quantum system interacting with electromagnetic field by the path integral formalism. We construct the amplitude and probability of quantum transition as path integrals in energy states space. The algorithm of path integral calculation was developed. This approach enables us to perform computer simulations of molecule dynamics induced by a laser field.

By the deduced formulas we describe quantum resonances in dynamics of nitrogen molecules, that interact with a sequence of ultrashort laser pulses.  The obtained results are in good agreement with the experimental data \cite{zhdanovich12} and the theoretical investigations \cite{floss13, floss12} by Schr\"odinger equation numerical solution.

The approach developed is appliable to nonperturbative studies of different multiphoton and nonresonant processes.

\begin{acknowledgments}
The work is supported by the Ministry of Education and Science of Russian Federation (grant 2.870.2011).
Numerical calculations were performed at Samara State Aerospace University by supercomputer "Sergey Korolev".
\end{acknowledgments}

\appendix

\section{Path integral formulation \\  in energy representation}
We consider the evolution operator kernel Eq.~(\ref{eq:9}) as a series and for the time interval $(t_k-t_{k-1})\rightarrow 0$ it is
\begin{eqnarray}
\langle l_k| \hat{U}_{D}(t_{k},t_{k-1})| l_{k-1}\rangle
=\langle l_k|l_{k-1}\rangle -
\nonumber \\
-\frac{\imath}{\hbar}\int\limits_{t_{k-1}}^{t_{k}}\langle l_k |\hat{V}_D(\tau)|l_{k-1}\rangle d\tau.
\label{eq:A1}
\end{eqnarray}

By using Eq.~(\ref{eq:2}) and Eq.~(\ref{eq:6}), the quantum system and electromagnetic field interaction operator is expressed as
\begin{equation}
\hat{V}_D(\tau)=
\sum\limits_{l',l=1}^{N}V_{l'l}(\tau)
\exp[\imath \omega_{l'l}\tau]  |l'\rangle \langle l|,
\label{eq:A2}
\end{equation}
where $V_{l'l}(\tau)$ -- interaction operator matrix element, $\omega_{l'l}=(E_{l'}-E_{l})/\hbar$ -- frequency of quantum transition between eigenstates with eigenvalues (energies) $E_{l'}$ and $E_{l}$.

Using Eq.~(\ref{eq:A2}), interaction operator matrix element in Dirac picture $\langle l_k |\hat{V}_D(\tau)|l_{k-1}\rangle$ is expressed
\begin{equation}
\langle l_k |\hat{V}_D(\tau)|l_{k-1}\rangle
=
V_{l_{k}l_{k-1}}(\tau) \exp[\imath \omega_{l_{k}l_{k-1}}\tau].
\label{eq:A3}
\end{equation}

Thus, we conclude
\begin{eqnarray}
\langle l_k| \hat{U}_{D}(t_{k},t_{k-1})| l_{k-1}\rangle
= \delta_{l_{k}l_{k-1}} - \nonumber
\\
-\frac{\imath}{\hbar}\int\limits_{t_{k-1}}^{t_{k}} V_{l_{k}l_{k-1}}(\tau) \exp[\imath \omega_{l_{k}l_{k-1}}\tau] d\tau.
\label{eq:A4}
\end{eqnarray}

Now we prove, that the kernel $\langle l_k| \hat{U}(t_{k},t_{k-1})| l_{k-1}\rangle$ of evolution operator can be expressed as
\begin{equation}
\langle l_k| \hat{U}_{D}(t_{k},t_{k-1})| l_{k-1}\rangle = \int\limits_{0}^{1} \exp[\imath S[l_k,l_{k-1};\xi_{k-1}]]d\xi_{k-1},
\label{eq:A5}
\end{equation}
where dimensionless action $S[l_k,l_{k-1};\xi_{k-1}]$ is found in Eq.~(\ref{eq:11}).

For this proof, by using Eq.~(\ref{eq:11}) we transform Eq.~(\ref{eq:A5}) into Eq.~(\ref{eq:A4}).
\begin{widetext}
\begin{eqnarray}
\langle l_k| \hat{U}_{D}(t_{k},t_{k-1})| l_{k-1}\rangle
=
\int\limits_{0}^{1} \exp[2\pi\imath(l_{k}-l_{k-1})\xi_{k-1}]\exp[-
\frac{\imath}{\hbar}\int\limits_{t_{k-1}}^{t_{k}} V_{l_{k}l_{k-1}}(\tau)
2\cos[2\pi (l_{k}-l_{k-1})\xi_{k-1} - \omega_{l_{k}l_{k-1}}\tau] d\tau]d\xi_{k-1} = \nonumber \\
=\int\limits_{0}^{1} \exp[2\pi\imath(l_{k}-l_{k-1})\xi_{k-1}]
(1 - \frac{\imath}{\hbar}\int\limits_{t_{k-1}}^{t_{k}} V_{l_{k}l_{k-1}}(\tau)
2\cos[2\pi (l_{k}-l_{k-1})\xi_{k-1} - \omega_{l_{k}l_{k-1}}\tau] d\tau)d\xi_{k-1}=
\nonumber \\
=
\int\limits_{0}^{1} \exp[2\pi\imath(l_{k}-l_{k-1})\xi_{k-1}]d\xi_{k-1} - \frac{\imath}{\hbar}\int\limits_{t_{k-1}}^{t_{k}} V_{l_{k}l_{k-1}}(\tau)
\int\limits_{0}^{1}
(\exp[4\pi\imath (l_{k}-l_{k-1})\xi_{k-1} - \imath\omega_{l_{k}l_{k-1}}\tau] +\exp[\imath\omega_{l_{k}l_{k-1}}\tau ])d\xi_{k-1}d\tau =
\nonumber
\end{eqnarray}
\end{widetext}
\begin{equation}
=\delta_{l_{k}l_{k-1}} - \frac{\imath}{\hbar}\int\limits_{t_{k-1}}^{t_{k}} V_{l_{k}l_{k-1}}(\tau)
\exp[\imath\omega_{l_{k}l_{k-1}}\tau]d\tau. \nonumber
\end{equation}

For this we use the facts, that the diagonal matrix element $V_{ll}(\tau)$ is equal to zero and integral representations of Kronecker symbol has the form:
\begin{equation}
\delta_{l_{k}l_{k-1}}=\int\limits_{0}^{1} \exp[2\pi\imath n(l_{k}-l_{k-1})\xi_{k-1}]d\xi_{k-1},
\label{eq:A6}
\end{equation}
where $n$ -- integer. 

So, we have proved the equivalence of Eq.~(\ref{eq:A5}) and Eq.~(\ref{eq:A4}), which define the quantum transition amplitude for time interval $t_{k}-t_{k-1} \rightarrow 0$. \vspace{1cm}

\section{Numerical simulation algorithm}

In this appendix we consider algorithm for numerical calculation of quantum transition amplitude $U(l_{f},t|l_{in},0)$ and probability $P(l_{f},t|l_{in},0)$.

The quantum transition amplitude calculation was made by recurrence relation
\begin{widetext}
\begin{equation}
\left(\begin{array}{c}
\Re[\tilde{U}(l_{k},t_{k}|l_{in},0)]\\
\Im[\tilde{U}(l_{k},t_{k}|l_{in},0)]

\end{array}\right)
=
\sum\limits_{l_{k-1}=0}^{N-1} \int\limits_{0}^{1}
\left(\begin{array}{c}
\cos[S[l_{k},l_{k-1};\xi_{k-1}]] \ \ -\sin[S[l_{k},l_{k-1};\xi_{k-1})]]\\
\sin[S[l_{k},l_{k-1};\xi_{k-1}]] \ \  \cos[S[l_{k},l_{k-1};\xi_{k-1})]]
\end{array}\right)
 \left(\begin{array}{c}
\Re[U(l_{k-1},t_{k-1}|l_{in},0)]\\
\Im[U(l_{k-1},t_{k-1}|l_{in},0)]
\end{array}\right)d\xi_{k-1},
\label{eq:B1}
\end{equation}
\end{widetext}
where  $\Re[\dots]$, $\Im[\dots]$ -- real and imaginary components; explicit form of $S[l_{k},l_{k-1};\xi_{k-1}]$ is defined by Eq.~(\ref{eq:11}).

The initial condition for pure quantum state $|l_{in}\rangle$ is as follows
\begin{eqnarray}
\left(\begin{array}{c}
\Re[U(l_{0},0|l_{in},0)]\\
\Im[U(l_{0},0|l_{in},0)]
\end{array}\right)
=
\left(\begin{array}{c}
\delta_{l_{0}l_{in}}\\
0
\end{array}\right).
\label{eq:B2}
\end{eqnarray}

Quantum transition probability $P(l_{k},t_{k}|l_{in},0)$ of investigated system from the state $|l_{in}\rangle$ at moment $t=0$ to the state $|l_{k}\rangle$ at moment $t_{k}$ can be expressed as

\begin{equation}
P(l_{k},t_{k}|l_{in},0) =
\Re[U(l_{k},t_{k}|l_{in},0)]^2 + \Im[U(l_{k},t_{k}|l_{in},0)]^2,
\label{eq:B3}
\end{equation}
where  normalized real and imaginary components of the transition amplitude  are

\begin{equation}
\left(\begin{array}{c}
\Re[U(l_{k},t_{k}|l_{in},0)]\\
\Im[U(l_{k},t_{k}|l_{in},0)]
\end{array}\right)
=
A^{-1}\left(\begin{array}{c}
\Re[\tilde{U}(l_{k},t_{k}|l_{in},0)]\\
\Im[\tilde{U}(l_{k},t_{k}|l_{in},0)]
\end{array}\right).
\label{eq:B4}
\end{equation}
The normalizing factor $A$ is calculated by the following formula:
\begin{equation}
A^2 = \sum\limits_{l_{k}=0}^{N-1} (\Re[\tilde{U}(l_{k},t_{k}|l_{in},0)]^2
+\Im[\tilde{U}(l_{k},t_{k}|l_{in},0)]^2).
\label{eq:B5}
\end{equation}

Using Eq.~(\ref{eq:B1})--(\ref{eq:B5}) we calculate  the amplitude $U(l_{f},t|l_{in},0)$ and probability $P(l_{f},t|l_{in},0)$ of the quantum transition for any $t$.

\nocite{*}

\bibliography{Shleenkov0714}

\end{document}